\documentclass[aps,prl,twocolumn,showpacs,preprintnumbers,eqsecnum,floatfix,nofootinbib]{revtex4}

\usepackage[dvips]{graphics,color,floatflt,epsfig} 

\usepackage{amsmath}

\usepackage{graphicx} 

\usepackage{psfrag}

\usepackage{bm}

\begin{document}
\DeclareGraphicsExtensions{.eps,.jpg, .mps, .png}
\preprint{\underline{FLO/4-QO}}
\newcommand{\Sref}[1]{Section~\ref{#1}}
\newcommand{\sref}[1]{Sec.~\ref{#1}}
\newcommand{\Cref}[1]{Chap.~\ref{#1}}
\newcommand{\tql}{\textquotedblleft} 
\newcommand{\tqr}{\textquotedblright~} 
\newcommand{\tqrc}{\textquotedblright} 
\newcommand{\Refe}[1]{Equation~(\ref{#1})}
\newcommand{\Refes}[1]{Equations~(\ref{#1})}
\newcommand{\fref}[1]{Fig.~\ref{#1}}
\newcommand{\frefs}[1]{Figs.~\ref{#1}}
\newcommand{\Fref}[1]{Figure~\ref{#1}}
\newcommand{\Frefs}[1]{Figures~\ref{#1}}
\newcommand{\reff}[1]{(\ref{#1})}
\newcommand{\refe}[1]{Eq.~(\ref{#1})}
\newcommand{\refes}[1]{Eqs.~(\ref{#1})}
\newcommand{\refi}[1]{Ineq.~(\ref{#1})}
\newcommand{\refis}[1]{Ineqs.~(\ref{#1})}
\newcommand{\framem}[1]{\overline{\overline{\underline{\underline{#1}}}}}
\newcommand{\PRA }{{\em Phys. Rev.} A }
\newcommand{\PRB }{{\em Phys. Rev.} B} 
\newcommand{\PRE }{{\em Phys. Rev.} E}
\newcommand{\PR}{{\em Phys. Rev.}} 
\newcommand{\APL }{{\em Appl. Phys. Lett.} }
\newcommand{\PRL}{Phys.\ Rev.\ Lett. }
\newcommand{\OCOM }{{\em Opt. Commun.} } 
\newcommand{\JOSA }{{\em J. Opt. Soc. Am.} A}
\newcommand{\JOSB }{{\em J. Opt. Soc. Am.} A}
\newcommand{\JMO }{{\em J. Mod. Opt.}}
\newcommand{\RMP}{Rev. \ Mod. \ Phys. }
\newcommand{\etal} {{\em et al.}}

\sloppy 
\thispagestyle{empty}

\def\ra{\rangle}
\def\la{\langle}
\def\bege{\begin{equation}}
\def\ende{\end{equation}}
\def\begarr{\begin{eqnarray}}
\def\endarr{\end{eqnarray}}
\def\no{\noindent}
\def\non{\nonumber}

\title{Single photons on demand from 3D photonic band-gap structures}

\author{Marian Florescu$^1$}
\email{marian.florescu@jpl.nasa.gov}
\author{Stefan Scheel$^2$}
\author{Hartmut H\"affner$^3$}
\author{Hwang Lee$^1$}
\author{Dmitry V. Strekalov$^1$}
\author{Peter L. Knight$^2$}
\author{Jonathan P. Dowling$^1$}
\affiliation{Jet Propulsion Laboratory, California Institute of
Technology, Pasadena, CA 91109-8099, USA\\
$^2$QOLS, Blackett Laboratory, Imperial College London, 
Prince Consort Road, London SW7 2BW, UK \\  
$^3$Institut f\"ur Experimentalphysik, Universit\"at
Innsbruck, Technikerstr. 25, A-6020 Innsbruck,  Austria}

\date{\today}

\begin{abstract}
We describe a practical implementation of a (semi-deterministic) photon gun based on stimulated
Raman adiabatic passage pumping and the strong enhancement of the photonic density of states in
a photonic band-gap material. We show that this device allows {\em deterministic} and {\em
unidirectional} production of single photons with a high repetition rate of the order of
100kHz. We also discuss specific 3D photonic microstructure architectures in which our model
can be realized and the feasibility of implementing such a device using $\mbox{Er}^{3+}$ ions
that produce single photons at the telecommunication wavelength of $1.55\mu$m.
\end{abstract}


\maketitle


In recent years, quantum optical information processing has attracted much attention, mostly
for its applications to secure communication protocols \cite{crypto} and the possibility of
solving efficiently computational tasks that are impossible to solve on a classical computer
\cite{computing}. Exploiting true single-photon sources --- rather than coherent pulses is a
goal of eavesdropper-proof quantum cryptography \cite{gilbert}. High-fidelity single-photon
sources are also a requirement for scalable linear optical quantum computing \cite{KLM}.

Present-day research considers photon emission from single atoms or molecules (either in cavity
QED \cite{Kuhn,varcoe00} or emission from color centers \cite{colourcentre}) quantum dot
structures \cite{QD}, mesoscopic p-n diode structures \cite{kim99}, chemical compounds
\cite{lounis00}, or micro-pillars \cite{pillars}.  Spontaneous parametric down-conversion, on
the other hand, may be used as a pseudo--single-photon source, conditioned upon detection of
one photon out of the pair \cite{pittman02}.  In the present proposal, we focus on the
possibility of modifying spontaneous emission by placing the radiation source inside a 3D
dielectric microstructure.  It is known that the rate of spontaneous decay of an excited atom
or ion can be tailored by the Purcell effect, whereby a cavity alters the density of modes of
the vacuum radiation field, which in turn can lead to enhancement or inhibition of spontaneous
decay of an atom inside the cavity \cite{Purcell46}.

\begin{figure}[htbp]
\centering\includegraphics[angle=0,width=0.9\linewidth]{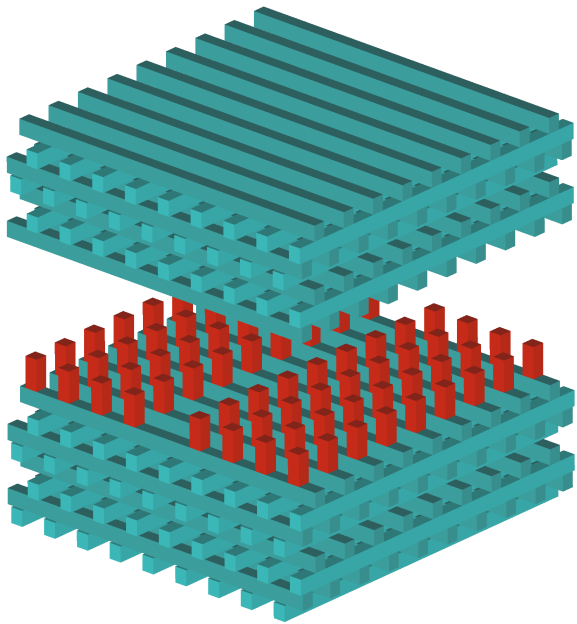}
\caption{ \label{chapter2_fig1} Photonic band-gap waveguide architecture for single-photon
generation. The micro-structure consists of a wave-guide channel in a 2D photonic crystal,
which is embedded in a 3D photonic crystal \cite{alpaper}. In this example, the 2D photonic
crystal consists of Si square rods of width $a_{2D}/a=0.3$ and thickness $h_{2D}/a=0.3$,
respectively (here $a$ is the dielectric lattice constant of the embedding 3D photonic
crystal). The linear waveguide is generated by removing one row of rods in the longitudinal
direction. The 3D photonic crystal is assumed to be a woodpile structure that presents a
photonic band gap of about 18\% of the mid-gap frequency \cite{alpaper}. The width and the
height of the stacking rods in the woodpile structure are $a_{3D}/a=0.25$ and $h_{3D}/a=0.3$,
respectively. }
\end{figure}

\begin{figure}[htbp]
\centering\includegraphics[angle=0,width=0.9\linewidth]{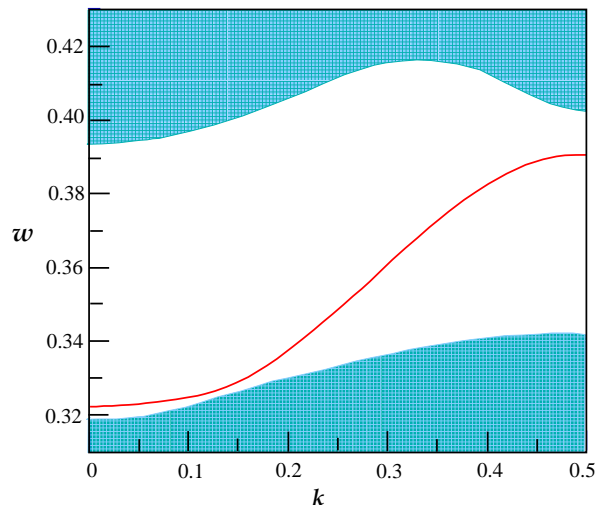}
\caption{ \label{chapter2_fig2} Schematic dispersion relation of a PBG heterostructure similar
to one presented in \fref{chapter2_fig1} for propagation along the waveguide direction (${\bm
w}= \omega a/2\pi c$, ${\bm k} = k_\parallel a/2\pi$).  By removing one row of rods, the linear
defect supports a single waveguided mode and, by appropriately choosing unit cell size, the
mode will experience a sharp cutoff in the spectral region around the $|2 \rangle \to
|1\rangle$ transition frequency.}
\end{figure}

Photonic crystals are periodically ordered dielectric materials that allow a very precise
control of the flow of light and of the light-matter interaction. In this paper, we make use of
a photonic crystal structures exhibiting a photonic band gap (PBG) \cite{pbgs}. In particular,
we focus on a simplified 1D model \cite{danieldos} of a complex 3D hetero-structure introduced
in \cite{alpaper}. In the context of single photon devices, PBG materials provide the
possibility of unidirectional enhancement of the atomic emission.  As shown in
\fref{chapter2_fig1}, a 1D photonic crystal model can be physically realized in a waveguide
channel in a 2D photonic crystal that is embedded in a 3D PBG material.  The electromagnetic
field is confined vertically by the PBG of the 3D structure (here, for example, we consider a
woodpile crystal \cite{woodpile}) and in-plane by the stop gap of the 2D photonic crystal (a
square lattice in this case) \cite{alpaper}. By tuning the characteristics of the
microstructure (geometry and index of refraction contrast) \cite{danieldos}, the linear defect
in the 3D PBG can support a single waveguide mode, which experiences a sharp cutoff in the gap
of 3D photonic crystal as shown in \fref{chapter2_fig2}. In this case, the sub-gap generated by
the waveguide channel has a true one-dimensional character, since there is only one direction
available for wave propagation. The sharp cutoff of the guided mode at the Brillouin zone
boundary gives rise to a low-group velocity ($d\omega/dk \to 0$), which combined with the
one-dimensional character of the system generates a divergent density of states (DOS)
($\rho(\omega) \propto dk/d\omega \to \infty$).  For an infinite structure, there is a physical
square-root singularity in the photonic density of states (DOS) near the cutoff of the
waveguide modes \cite{quang95}.
For a finite structure, the divergence is removed by the finite-size effects
\cite{Bendickson96}. However, the strong variation with frequency of the photonic DOS remains
\cite{danieldos}. Therefore, while we limit the present analysis to an idealized 1D photonic
crystal in which the single-mode waveguide channel is modeled as an {\em effective} 1D photonic
crystal consisting of alternating double-layer of quarter wave plates, we emphasize that this
model can be implemented in carefully designed 3D dielectric hetero-structures (the
electromagnetic field in the guided mode encounters a periodic 1D effective variation of the
dielectric constant as it propagates along the waveguide channel).  The most relevant features
of the photonic crystals for realization of single photon \tql gun\tqr devices (the rapid
variation with frequency of the DOS and the unidirectional operation) are easily recaptured in
the simplified 1D model. Essentially, for an $N=29$ period stack with an index of refraction
contrast of 2:1, the spontaneous decay rate at the band edge frequency is enhanced by a factor
of 115 compared to inside the bulk dielectric.  As the number of periods $N$, becomes larger,
the asymptotic behavior of the density of mode is given by $\rho^{\rm BER}_{N}\!  \propto N^2
\!\rho^{\rm bulk}$, where $\rho^{\rm BER}_{N}$ and $\rho^{\rm bulk}$ represent the density of
modes at band-edge resonance and in bulk \cite{Bendickson96}.

We emphasize that only in one-dimensional systems the low-group velocity modes give rise to
variations with the frequency of the optical DOS strong enough to drive a \tql on-demand\tqr
emission of photons. In conventional 3D photonic crystals, the contribution of the low-group
velocity modes to the DOS is of an integrable form and the DOS, while presenting
discontinuities of the slope, $d \rho/d\omega$, remains finite and continuous as function of
frequency \cite{florescu1}. There are physical systems in which these one-dimensional models
can be practically implemented.

\begin{figure}[ht]
\includegraphics[width=0.6\linewidth]{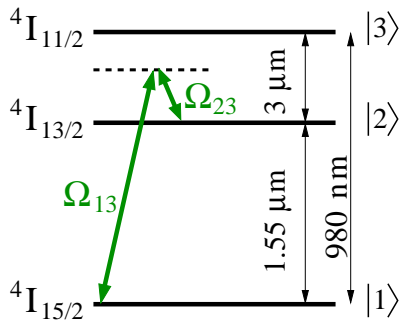}
\caption{\label{fig:er} Schematic level diagram of Er$^{3+}$ ion discussed in the context of
the deterministic STIRAP pumping scheme.  Using the technique of STIRAP, the population
transfer to the level $|2\ra$ does not depend on the branching ratio and can be made with unit
efficiency.  }
\end{figure}

We now consider an $\mbox{Er}^{3+}$ ion embedded in the dielectric backbone of the PBG. The ion
could be placed with an atomic-force microscope or by sparse ion implantation midway during the
structure's growth \cite{clarks}. The wavelength of $1.55\mu$m of its ${}^4\mbox{I}_{13/2}
\rightarrow {}^4\mbox{I}_{15/2}$ transition ($|2\ra \rightarrow |1\ra$, for short) is most
convenient for quantum communication with optical fibers \cite{desurvire94}.  Excitation can be
performed by pumping at $980$ nm, corresponding to the ${}^4\mbox{I}_{11/2}
\rightarrow{}^4\mbox{I}_{15/2}$ transition ($|3\ra \rightarrow |1\ra$), see Fig.~\ref{fig:er}.
However, even a 100\% efficient population transfer to the upper state $|3\ra$ does not
guarantee that it will decay down to the state $|2\ra$.  For example, the level
${}^4\mbox{I}_{11/2}$ decays to ${}^4\mbox{I}_{15/2}$ about six times faster than to the level
${}^4\mbox{I}_{13/2}$ \cite{dierolf99}.  Instead, in-band pumping at 1480 nm may be used
\cite{baumann96}. Nevertheless, due to the relatively small radiative decay rate of the $|2
\rangle \to |1\rangle$ transition, in-band pumping is not very efficient and, more importantly,
does not have a deterministic character.

A more efficient and {\em deterministic} preparation of the emission-ready state $|2\ra$ of the
$\mbox{Er}^{3+}$ ion can be carried out by using stimulated Raman adiabatic passage (STIRAP)
from the ground state \cite{bergmann98}.  This method allows, in principle, for a 100\%
population transfer even for a strongly decaying intermediate state $|3\ra$.  A {\em
deterministic} population transfer is achieved by first turning on the $\Omega_{23}$, and then
turning on the $\Omega_{13}$ while $\Omega_{23}$ is turned off, all done adiabatically.  Here
$\Omega_{ij}$ denotes the Rabi frequencies of the STIRAP fields coupling the levels $|i\ra$ and
$|j\ra$.  If both pulses have the same Gaussian shape, the most efficient transfer is achieved
when their peaks are separated by about $\tau$, and the adiabaticity condition $\tau
\sqrt{\Omega_{13}^2 + \Omega_{23}^2 }>10$ is fulfilled \cite{bergmann98}.  For atoms with
dipole transitions in the IR optical range, the adiabaticity and optimization requirements are
easily satisfied for relatively low-power, nanosecond pulses with nearly transform-limited
spectral width.

The STIRAP process provides both the pump and the trigger mechanisms of the single-photon-gun
device.  After the {\em deterministic} excitation process, the ion is left in its $|2 \ra$
state for a time that is inversely proportional the spontaneous decay rate of the metastable
state. Assume now we could arrange the properties of the dielectric microstructure in such a
way that the transition $|2\ra \to |1\ra$ frequency is placed in the spectral region
surrounding the cutoff frequency of the waveguide mode. After the excitation process, the ion
will feel a large density of modes and will decay very rapidly.  This type of process we call
\tql on demand\tqr since the onset of spontaneous decay can be controlled externally.  The
process becomes more deterministic the higher the local density of modes gets, that is, the
sharper the band edge becomes.  This increase in mode density can be achieved by increasing the
longitudinal size (number of periods) of the photonic crystal.

We also note that photonic crystal heterostructure architecture in \fref{chapter2_fig1} has an
additional advantage for practical implementations of a single-photon-gun device. By increasing
the transverse size of the waveguide channel, the linear defect may support additional guided
modes.
These additional modes can be used to convey the external laser fields that drive the pumping
process.

\begin{figure}[ht]
 \includegraphics[width=0.9\linewidth]{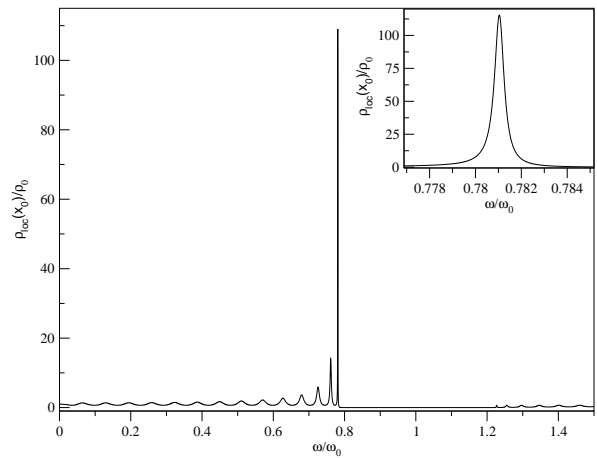}
\caption{\label{fig:se} Normalized local mode density at the ion position for a 29-layer stack
with $n_1=1$, $n_2=2$. The ion is placed in the middle of the dielectric slab situated halfway
between the longitudinal boundaries of the structure. The inset shows an expanded view of the
spectral region surrounding the band edge frequency. The line-width of the Er$^+$ ion itself is
about 0.0001 in these dimensionless units \cite{Erion}.}
\end{figure}

For concreteness, consider the model where-bye the radiating ion is placed in the middle of an
effective 29-layer dielectric structure with $n_1=1$ and $n_2=2$. The value of $n_2=2$ is
somewhat arbitrary.  The woodpile structure can be made out of very high index of refraction
materials in an air matrix, such as Si or III-IV semiconductors, GaAs or InP etc. This means
that we could use in our simulations a larger $n_2$ [$n_2 \in (3.14, 3,5)$] index of refraction
contrast. However, we employ an effective 1D model, and we expect the effective index of
refraction to be somewhat lower than the actual index of refraction of the dielectric backbone.
Suppose the ion is in the excited state $|2\ra$ and the transition frequency $\omega_A$
corresponds to $0.781 \omega_0$ (here $\omega_0$ is the mid-gap frequency).  Figure
\ref{fig:se} shows the normalized spontaneous emission rate (with respect to the low frequency
emission rate) that is proportional to the local DOS at the ion position (in this example, the
ion is placed in the middle dielectric slab of the {\em effective} one dimensional photonic
crystal).  Note that due to the effective one-dimensional character of the device, the
enhancement of the mode density preferentially occurs at a single mode of propagation.
Correspondingly, the emission is highly directional along the waveguide channel and may be
easily mode-matched to, say, a telecom fiber or other waveguide.

We now consider the issues related to the speed of such a photon gun.  The maximum repetition
rate of that device is limited by the following factors: (1) the repetition rate of the ion
excitation, and (2) the spontaneous decay rate (inverse lifetime) of the metastable ion
state. For mode-locked laser diodes the repetition rate can be as high as 1 GHz, but is
certainly in the 100 MHz range.  That means that the limiting factor is the enhanced
spontaneous decay rate of the erbium ion.  Taking the lifetime of the excited ion to be 1 ms in
the bulk \cite{desurvire94}, the enhanced spontaneous decay rate of the erbium ion can be of
the order of 100 kHz, using the model above.  But as mentioned earlier, this rate can be
increased by sharpening the band edge with the addition of more periods or by increasing the
index contrast.  The increase of the mode density scales as $N^2$, where $N$ is the number of
the periods.  The total repetition rate of the device then can be as high as several MHz for
realistic $N$.

We also note that our single-photon device proposal eliminates the inverse relationship between
the magnitude of the defect-mode DOS and the repetition rate of the device that would be
present in a 3D PBG single defect-mode based proposal for a single photon gun (along the lines
of Ref.~\cite{others}). In such a defect-mode device, the repetition rate is unfavorably
limited by the cavity build-up time, which is inversely proportional to the quality factor of
the defect mode, whereas the enhancement of the DOS is proportional to the the quality factor
of the defect mode. Moreover, while the Purcell enhancement and emission narrowing effects in a
1D PBG structure rely crucially on preparing the emitting Er$^{3+}$ dipole oriented parallel to
the stack interfaces \cite{dowling93}, this is not the case for 3D PBG structures, where mode
suppression and enhancement are omni-directional, independent of dipole orientation.

An alternative option for triggering the single photon emission process would be to make use of
an intensity-dependent Kerr nonlinearity embedded in the dielectric backbone of the photonic
crystal. If initially the ion frequency $\omega_A$ falls inside the photonic band gap, the ion
can not decay due to the lack of photonic modes. By applying an external optical or electric
field (that induces a prescribed change of the refraction index of the nonlinear material and
shifts the band gap to a different frequency interval), the ion transition frequency will now
be located within the continuum of modes near the photonic band edge, and will suddenly feel a
strong DOS and will spontaneously decay very rapidly. Our calculations show that for a
dielectric structure consisting of 39 unit cells the required nonlinear relative change of the
refraction index necessary to achieve single photon generation processes is $\Delta n/n\approx
6\times10^{-3}$. Moreover, using the photonic crystal architecture presented in
\fref{chapter2_fig1}, we argue that the external laser field power required to achieve the
necessary change in the index of refraction may be strongly reduced.  By engineering the
symmetry of the field distribution in the photonic crystal, one may achieve very strong field
local enhancement at the nonlinear medium location, and, implicitly, a strong variation of the
nonlinear index of refraction with only a fraction of the power that would have been required
to obtain the same variation in the case of a homogeneous medium.

 In summary, we have proposed a source of single photons switched by a STIRAP pumping process
of a mono-atomic source placed in a light confining dielectric structure.  The atomic source
can be rapidly switched, at will, with a high repetition rate. The virtue of the mode
confinement effect in the architecture presented in \fref{chapter2_fig1} is that the single
photon gun device can be made small and compact. The unidirectional operation of the single
photon device is achieved by tailoring the PBG geometry, while the repetition rate of the
device is dramatically increased due to the strong enhancement of the optical DOS near a
photonic band edge.

This work was partially supported by the Feodor-Lynen program of the A.~v.~Humboldt foundation
(SS) and the QUEST program of the European Union (HPRN-CT-2000-00121).  Part of this work was
carried out at the Jet Propulsion Laboratory, California Institute of Technology, under a
contract with NASA. We would like to acknowledge support from NSA, ARDA, DARPA, NRO, ONR,
respectively. MF and HL would like to acknowledge the NRC, as well as NASA Codes S and Y, for
additional support.


\end{document}